\documentclass[a4paper,UKenglish,cleveref,thm-restate]{lipics-v2021}

\nolinenumbers{}

\newcommand{\parag}[1]{\noindent\textbf{\textsf{#1}}}
\newcommand{\itemname}[1]{\textbf{\textsf{#1}}}

\newcommand{\rank}{\mathsf{rank}}
\newcommand{\rankone}{\mathsf{rank}}
\newcommand{\rankall}{\mathsf{rank_4}}

\usepackage{ragged2e} % to justify text in abstract
\usepackage{graphicx}
\usepackage{svg}
\usepackage{balance}
\usepackage{xcolor}
\usepackage{xspace}
\usepackage[utf8]{inputenc}
\usepackage{float}
\usepackage{amssymb}
\usepackage{amsmath}
\usepackage{multirow}
\usepackage{booktabs}
\usepackage{makecell}
\usepackage{amsthm}
\usepackage{thmtools}
\usepackage[makeroom]{cancel}
\usepackage{xcolor, soul}
\usepackage{siunitx}
\usepackage{tablefootnote}

\usepackage{fancyvrb}
\VerbatimFootnotes

\usepackage[noEnd=true,indLines=true]{algpseudocodex}
\usepackage[ruled]{algorithm}
\tikzset{algpxIndentLine/.style={draw=black}}

\usepackage{afterpage}

\hypersetup{
    colorlinks=true,
    linkcolor=blue,
    citecolor=blue,
    filecolor=magenta,
    urlcolor=magenta
}

\usepackage[newfloat,frozencache,cachedir=.]{minted}

\newmintedfile{rust}{
  numbersep=5pt,
  frame=lines,
  baselinestretch=1.05,
  fontsize=\footnotesize,
}

\title{QuadRank: Engineering a High Throughput Rank}

\author{Ragnar Groot Koerkamp}{Karlsruhe Institute of Technology,
  Germany}{ragnar.grootkoerkamp@gmail.com}{https://orcid.org/0000-0002-2091-1237}{}

\authorrunning{R. Groot Koerkamp}
\Copyright{Ragnar Groot Koerkamp}

\supplementdetails[cite={}, swhid={swh:1:dir:83a2ce8d57ab4f246e470f1226e19ecafb51fffe}]{Software}{https://github.com/RagnarGrootKoerkamp/quadrank}

\acknowledgements{
I thank Heng Li for motivating me to start this project in an attempt to
speed up BWA-MEM. I also thank those who were involved in discussions
surrounding this project or gave feedback on the text:
Rick Beeloo, Piotr Beling, Felix Leander Droop, Simon Gene Gottlieb, Florian Kurpicz,
Rob Patro, Giulio Ermanno Pibiri, and Peter Sanders.
}


\hideLIPIcs

\begin{document}

\hypersetup{pageanchor=false}
\maketitle

\begin{abstract}\label{sec:orgeb7ba07}%
\parag{Motivation.} Given a text,
a query \(\rank(q, c)\) counts the number of occurrences of
character \(c\) among the first \(q\) characters of the text.
Space-efficient methods to answer these rank queries form an important
building block in many succinct data structures.
For example, the FM-index \cite{fm-index} is a widely used data
structure that uses rank queries to locate all occurrences of a pattern in a text.

In bioinformatics applications, the goal is usually to process large inputs as
fast as possible. Thus, data structures should have high \emph{throughput} when used
with \emph{many threads}.

\parag{Contributions.}
We first survey existing results on rank data structures.
For the \(\sigma=2\) binary alphabet, we then develop BiRank, which has 3.28\% space overhead.
BiRank merges the central ideas of two recent papers: (1)
we interleave (inline) offsets in each cache line of the underlying bit vector
\cite{spider}, reducing cache misses, and (2)
these offsets are to the \emph{middle} of each block so that only half of each needs
popcounting \cite{engineering-rank}.
In QuadRank (14.4\% overhead), we extend these techniques to the \(\sigma=4\) (DNA) alphabet.

Both data structures typically require only a single cache miss per query, making
them highly suitable for high-throughput and memory-bound settings.
To enable efficient batch-processing, we support \emph{prefetching} the cache lines required to answer upcoming queries.

\parag{Results.}
BiRank and QuadRank are around 1.5\(\times\) and 2\(\times\) faster than
similar-overhead methods that do not use interleaving.
Prefetching gives an additional 2\(\times\) speedup, at which point the
dual-channel DDR4 RAM
bandwidth becomes a hard limit on the total throughput. With prefetching, both
methods outperform all other methods apart from SPIDER \cite{spider} by 2\(\times\).

When using QuadRank with prefetching in a toy count-only FM-index, QuadFm, this
results in a smaller size and up to 4\(\times\) speedup over Genedex, a state-of-the-art batching
FM-index implementation.

\parag{Conclusion.}
Optimizing data structures for high throughput, by minimizing
cache misses and branch-misses and adding support for prefetching, can result in
significant speedups when benchmarks are adjusted accordingly.
\end{abstract}

\newpage

\hypersetup{pageanchor=true}
\setcounter{page}{1}

\section{Introduction}
\label{sec:org5fa6675}
Given a fixed text \(T=t_0\dots t_{n-1}\) of length \(n\) over an alphabet \(\Sigma\) of size \(\sigma\), a
query \(\rank(q, c)\) counts the number of occurrences of symbol \(c\in
\Sigma\) in the first \(q\) (\(0\leq q\leq n\)) characters of the text\footnote{Like
Rank9 \cite{rank9} and most (but not all) other implementations, we follow Dijkstra's advice \cite{dijkstra-numbering} and
start numbering at zero.}:
\vspace{-0.3em}
$$
\rank(q, c) := \sum_{i\in \{0, \dots, q-1\}} [t_i = c].
$$
In most literature, the binary alphabet of size \(\sigma=2\) is used, in which
case the text is simply a string of \(n\) bits. In this case, we also write
\(\rank(q) := \rank(q, 1)\) to count the number of \(1\) bits.

Of interest are space-efficient data structures that
can answer these queries quickly. Indeed, there exist \emph{succinct} data structures
\cite{succinct-data-structures} that use \(n + o(n)\) bits of space to answer
queries on a binary text in \(O(1)\) time in the RAM-model with word-size
\(w=\Theta(\lg n)\). 
When the bitvector itself is stored explicitly,
a tight lower bound on the space usage is \(n + \Omega(n \log\log n / \log n)\)
bits \cite{rank-space-bound,rank-optimal-space-bound}.

A fast and widely used implementation is Rank9 \cite{rank9}, which has a fixed \(25\%\) space overhead.
Many subsequent works have reduced the space overhead to as little as 1.6\%, as detailed in \cref{sec:org94c5378}.
In practice, most implementations have fixed overhead, making them \emph{compact} (\(n+O(n)\) bits) but not \emph{succinct}.

\parag{FM-index.}
A primary application of Rank queries is in the \emph{FM-index} \cite{fm-index}, a
succinct data structure that can efficiently locate all occurrences of a pattern in a
text and is used in tools such as BWA-MEM \cite{bwa-mem}, and Bowtie
\cite{bowtie,bowtie2}.
Whereas most of the literature on rank structures assumes a binary
alphabet (\(\sigma=2\)), in this case the DNA alphabet has size \(\sigma=4\).
Indeed, BWA-MEM implements its own rank structure over a 2-bit alphabet\footnote{\url{https://github.com/lh3/bwa/blob/master/bwt.c}},
and this paper started as an attempt to speed this up.

\parag{Wavelet tree.}
For alphabets of arbitrary size, \emph{wavelet trees} \cite{wavelet-tree}
or the \emph{wavelet matrix} \cite{wavelet-matrix}
can be used for succinct rank queries.
They both need \(\lg_2 \sigma\) queries to a binary rank structure. Recently,
quad wavelet trees \cite{quad-wavelet-tree} have been introduced, following
earlier theoretical \cite{compressed-representations} and practical
\cite{multiary-wavelet-trees} results on multi-ary wavelet trees.
Quad wavelet trees use rank over a \emph{quad vector} as a building block,
and thus need only \(\log_4 \sigma\) rank queries, leading to 2\(\times\) to
3\(\times\) speedups over binary wavelet trees.

\parag{Multithreading and batching.}
In the past, increasing CPU frequencies led to faster code, but nowadays,
improvements are mostly in increasing parallelism. Furthermore, total compute of
a CPU increases faster than the available memory bandwidth, resulting in the
need for \emph{communication-avoiding algorithms} that minimize their memory
bandwidth \cite{evolution-of-mathematical-software}.
Indeed, in bioinformatics applications, one often has many independent queries (DNA sequences)
that need to be processed (searched in an FM-index) as fast as possible.
In particular, this allows using all cores/threads of the CPU as well as
processing queries in \emph{batches} inside each thread, to hide the memory latency.

Current benchmarks usually measure the throughput of answering rank queries in a
for loop on a single thread, but this does not take into account the possibility for batching, nor
does it capture the effects of running many threads in parallel.
As we will see, in a high-throughput setting, many existing methods become
bottlenecked by the total memory bandwidth of the CPU. We specifically
design our data structures to use the memory bandwidth maximally efficient.

\parag{Contributions.}
We develop two data structures, BiRank and QuadRank, that support
high-throughput rank queries over texts over alphabets of size 2 and 4.

Both of them integrate a number of existing techniques (see next section),
and are \emph{not} designed to support select queries, since these are not
needed for the motivating FM-index application, thus allowing for more optimizations.
Specifically, BiRank has 3.28\% overhead and integrates (1) inlining of counts into the bitvector
\cite{spider}, which reduces cache misses, (2) 
\emph{pairing} with mask-lookup \cite{engineering-rank}, halving the number of popcounts,
and (3) an additional third layer \cite{poppy} that is modified to be
only half its usual size.

QuadRank extends the ideas of BiRank, but has roughly 4\(\times\) larger space
overhead (14.4\%)
since it stores metadata for each symbol. It combines the cache locality of the
implementation in BWA-MEM \cite{bwa-mem} with the low overhead of
quad vectors \cite{quad-wavelet-tree-preprint} and a transposed bit layout for
faster queries \cite{awry-optimized-fm-index,engineering-rank}.
QuadRank is optimized for returning ranks for all 4 symbols at once by using
AVX2 instructions, which is useful for approximate pattern matching in an FM-index.

Both data structures need only a single cache line from RAM to answer
queries as long as the input is less than roughly 16 GiB.
The main novelty is in combining all these existing ideas and applying them to
the size-4 alphabet, while also adding support for \emph{prefetching} of
cache lines to enable much more efficient batch-processing of queries.
As a side-effect, we also added prefetching to some other libraries, yielding up
to 2\(\times\) speedups.

\parag{Results.}
For both data structures, we implement a number of variants that have different
space-time trade-offs.
When used in a for loop, BiRank is up to 1.5\(\times\) faster than the next-fastest rust
implementation of equal size, with the speedup being larger when using many
threads.
Prefetching memory improves the throughput of many libraries by around 1.5\(\times\), and
improves BiRank by 2\(\times\). In this setting, all methods are bottlenecked by
the memory throughput, and BiRank is 2\(\times\) faster than all others because it
only needs to read 1 instead of 2 cache lines from RAM.
Similarly, QuadRank is at least 1.5\(\times\) faster than the next-fastest Rust
library, QWT \cite{quad-wavelet-tree}, and 2\(\times\) faster after adding
prefetch instructions, again being bottlenecked by the RAM throughput.

Inspired by genedex \cite{genedex}, we develop QuadFm, a
toy-implementation of a count-only FM-index that uses batching, prefetching, and
multithreading. At 14.4\% overhead (2.29 bits/bp), our implementation is over 1.5\(\times\) faster 
using QuadRank compared to using QWT's quad vector,
and at 100\% space overhead, QuadFm is 4\(\times\) faster than genedex, a
state-of-the-art FM-index implementation.

\section{Background}
\label{sec:org94c5378}
We briefly go over some previous papers containing rank structures for either
\(\sigma=2\) or \(\sigma=4\) in chronological order and list their main technical contributions.
Both \cite{poppy} and \cite{pasta} contain a nice
overview as well.
Note that many of these papers develop a rank structure in the context of the
larger \emph{rank and select} problem, where there are different design trade-offs.
Additionally, work on \emph{compressed} bitvectors is omitted here.

\begin{figure}[htbp]
\centering
\makebox[\textwidth][c]{
\includesvg[width=1.15\linewidth]{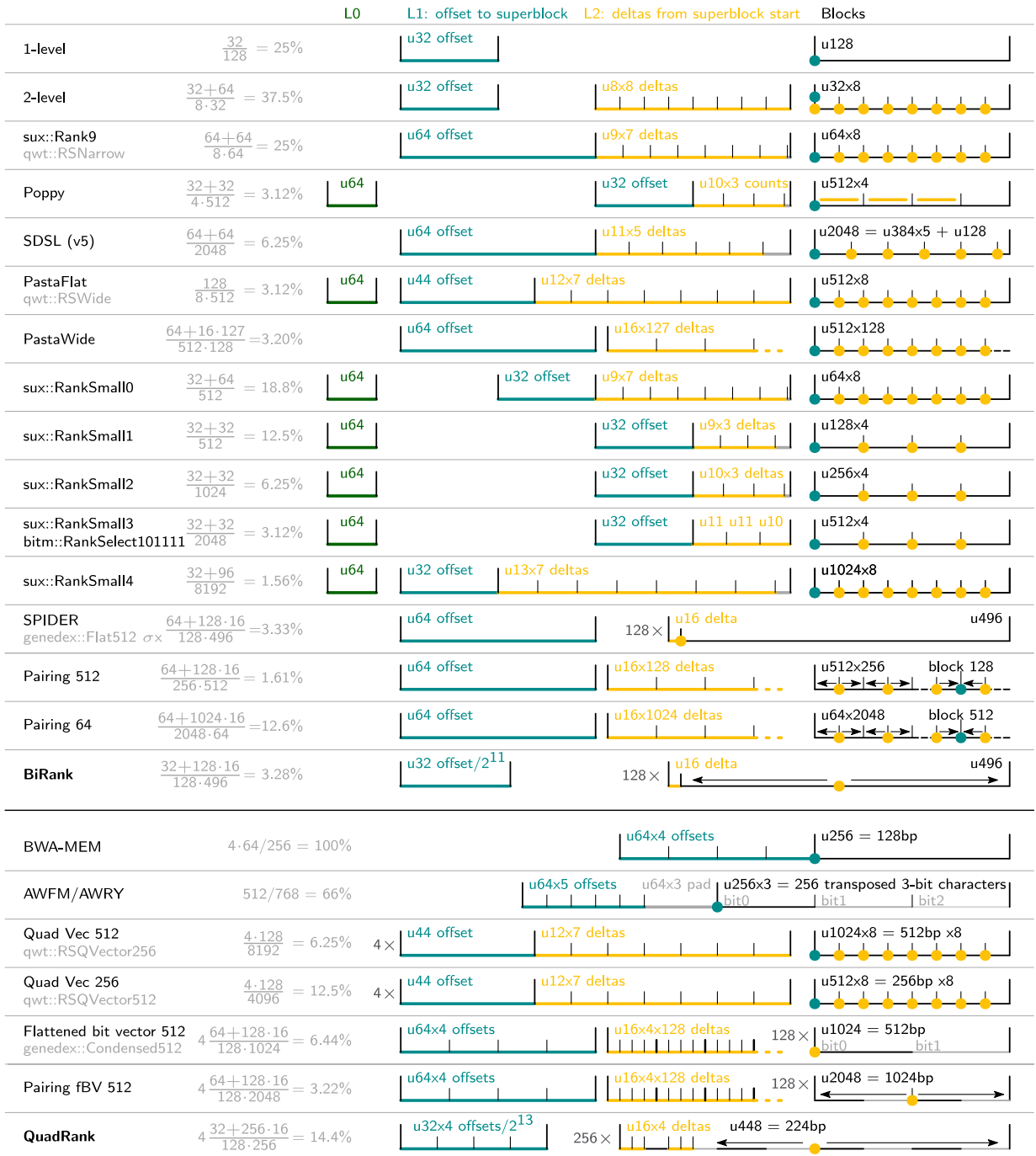}
}
\caption{\label{ranks}Schematic overview of rank data structures. The top and bottom half are for \(\sigma=2\) and \(\sigma=4\) respectively. Each line shows a data structure (and notable (re)implementations) with its overhead and the layout of a single \emph{superblock} (not to scale). Each structure stores up to 3 vectors containing (interleaved) superblocks offsets, block deltas, and raw bits. On the right (black) are the \emph{blocks} containing (bitpacked) data. Each superblock contains a single L2 \emph{offset} (teal) that is either absolute, or sometimes relative to a 64-bit L3 value (green). They usually count the number of 1-bits/characters before the start of the superblock as indicated by the teal dot, or to the middle of the superblock for pairing variants. L1 \emph{deltas} (yellow) count from the start/middle of the superblock to the start of each block (yellow dots). Only for poppy they count individual blocks (yellow lines). For pairing, pairing fBV, BiRank, and QuadRank, L1 deltas are to the middle of each (pair of) block(s). AWFM, (pairing) fBV, and QuadRank store the text \emph{transposed}, alternating words of low and high bits.}
\end{figure}

Most data structures are schematically depicted in \cref{ranks}.

\parag{Terminology.}
For later reference, we summarize our terminology.
The raw data is split into \emph{superblocks} (the second level, L2) that are further
split into \emph{blocks} (L1). \emph{Block} is used for both the raw bits themselves, as well as the
cache line containing them.
For each superblock, an L2 \emph{offset} is stored, representing the number of 1-bits
before the superblock. For each block, an L1 \emph{delta} is stored, typically representing the
number of 1-bits preceding it inside the superblock.
Conceptually, these levels form a \emph{summary tree}, named as such by
Kurpicz et al. \cite{rank-select-theory-practice}. We follow their notation,
and number the levels of the tree \emph{bottom-up}, breaking tradition with e.g. the
presentation in \cite{poppy} and \cite{pasta}.
The root has all superblocks (L2) as children,
each superblock has its contained blocks (L1) as children,
and each block has the contained bits as children.

The \emph{overhead} of a data structure is the increase in space consumption relative
to the size of the input data.
We use bp (base pair) as the unit for 2-bit encoded DNA characters, and
occasionally use the Rust syntax \texttt{u64} for a 64-bit variable and \texttt{u64x4} for a
256-bit SIMD register containing 4 64-bit words.
A \emph{symbol} is an element of the alphabet \(\Sigma\), whereas
a \emph{character} is an element of a string.

\parag{Classic succinct approach.}
As a baseline, Jacobson \cite{succinct-data-structures}
stores the bitvector, and additionally two levels of blocks alongside this.
\emph{Blocks} consist of \(\lfloor\log(n)/2\rfloor\) bits,
and \(\lfloor\log n\rfloor\) blocks form a \emph{superblock}.
Level L2 of the tree then contains a \(\lceil \log n\rceil\) bit \emph{offset} for each
superblock, counting the number of set bits preceding it.
Level L1 stores for each block a \(\lceil\log \log n\rceil\) bit
\emph{delta} counting the number of 1-bits preceding it inside its superblock.
The number of 1-bits in a (prefix of) a block is obtained via a lookup in a
precomputed table of size \(2^{(\log n)/2} = \sqrt n\).

\parag{A practical approach.}
González et al. \cite{practical-rank-select} observe that the classic method above
has 66.85\% overhead in practice for \(n=2^{30}\).
They replace the large lookup table by a smaller table of
per-byte popcounts. (Meanwhile, CPUs natively support 64-bit \texttt{popcount}
instructions.)
They use 256-bit superblocks with a 32-bit offset, containing 8
32-bit blocks, each with their own 8-bit delta.
Alternatively, they introduce a \emph{single}-level tree
storing a 32-bit L2 offset after every e.g. \(4\cdot 32\) bits and omitting L1.
This requires popcounting more words, but has the benefit of improved
cache locality compared to a two-level tree.

\parag{Rank9: interleaving levels.}
\emph{Rank9} \cite{rank9} has 25\% overhead and \emph{interleaves} the L2 and L1 levels of the classic tree.
Each block is 64
bits, and 8 blocks form a 512-bit superblock, exactly matching a cache line.
For each superblock, the interleaved
tree stores a 64-bit integer with the offset of the superblock, and 7
9-bit deltas (for all but the first block) in an additional 64-bit word.
This needs two cache misses per query (for the L2 array and bits), and is very fast in practice.
Specifically it only needs to popcount a single 64-bit word, which is done
using \emph{broadword programming} (also known as SWAR, SIMD Within A Register).

\parag{Poppy: reducing space.}
\emph{Poppy} \cite{poppy} is optimized for space and has only 3.125\% overhead.
First, it makes the observation that performance is largely determined by the
number of cache misses. Thus, it uses larger blocks of 512 bits. It then re-uses
Rank9's interleaved index with two modifications.
Each superblocks contains 4 blocks, and for each superblock it stores a 32-bit
offset (L2) followed by 3 10-bit popcounts of the first 3 blocks.
Queries then require a prefix-sum over these counts.
To handle 64-bit outputs, it stores an additional layer (L3) of the tree,
with a full 64 bit offset after every \(2^{32}\) input bits.

\parag{BWA-MEM: DNA alphabet.}
BWA-MEM \cite{bwa-mem} implements a 100\% overhead rank data structure on \(\sigma=4\) DNA.
It interleaves L2 offsets with the data, and requires only a single cache miss per query.
In each cache line, it stores 4 64-bit offsets (one for each DNA character),
followed by 256 bits encoding 128 bp.

\parag{SDSL.} The succinct data structure library (SDSL) \cite{sdsl} implements Rank9
and introduces \texttt{rank\_support\_v5}, which has 6.25\% overhead. It uses superblocks of
2048 bits. For each, it interleaves a 64-bit offset (L2) and 5 11-bit deltas (packed
into 64 bits) to all but the first of 6 blocks covering \(6\cdot 64\) bit.
\texttt{rank\_support\_il} interleaves 64-bit offsets with 512-bit blocks.

\parag{EPR-dictionaries: arbitrary \(\sigma\).}
EPR-dictionaries \cite{epr-dictionaries} work for arbitrary alphabet. For
\(\sigma=4\), they use 64-bit (32 bp) blocks and have 42\% overhead, and
interleave an independent 2-level rank structure for each character.
Compared to earlier work, space is saved by storing a packed representation
of the text instead of \(\sigma\) (1-hot) encoded bitvectors that each indicate
which text positions contain each symbol \(c\in \Sigma\).

\parag{B-trees.}
Pibiri and Kanda \cite{rank-select-mutable-bitmaps} diverge from the classic approach and
introduce a rank and select structure based
on highly tuned B-trees that have 3.6\% overhead. Each rank query traverses
roughly \(\log_{16} n\) levels of the tree, with the middle levels packing 16
32-bit values in a cache line. Due to efficient caching of the top levels of the
tree, performance is similar to poppy, although not as fast as rank9.

\parag{AWFM: transposed layout and batching/prefetching.}
The AWFM-index and its Rust implementation AWRY \cite{awry-optimized-fm-index} builds an FM-index on a size \(\sigma=6\)
alphabet of 4 DNA characters as well as a sentinel and ambiguity symbol.
It uses blocks of 256 3-bit characters, preceded by 5 64-bit offsets that are padded to
512 bits. Each block is encoded using a similar \emph{strided} or \emph{transposed layout}:
instead of concatenating the 3 bits of each character, it stores 3 256-bit
vectors containing bit 0, bit 1, and bit 2 of each character.
This allows for more efficient popcounting.
The FM-index processes queries in batches of size 4, and prefetches memory
needed for the next rank operation as soon as possible.

\parag{Pasta: larger L1 values and faster queries.}
\emph{PastaFlat} \cite{pasta,pasta-preprint} has the same 3.125\% space overhead as Poppy,
but improves query time by 8\% by  avoiding Poppy's need to take a prefix
sum over L1 counts. Pasta doubles the metadata for each superblock to 128 bits, covering
8 512-bit blocks of 4096 bits in total. It stores a 44-bit offset (L2) followed by 7 12-bit deltas (L1) from the start of the
superblock to each block.
A second structure, \emph{PastaWide} (3.198\% overhead) uses 16-bit values for L1, which allows faster
select queries using SIMD instructions.
Here, each superblock covers 128 blocks and stores a 64-bit L2 value, this time \emph{not}
interleaved with the L1 values, and the L3 level is dropped.

\parag{Quad vectors: extending PastaFlat to \(\sigma=4\).}
\emph{Quad wavelet trees} internally use \emph{quad vectors}
\cite{quad-wavelet-tree-preprint,quad-wavelet-tree}, which have a layout very
similar to PastaFlat.
Super blocks cover eight 512 bp blocks and stores 128 bits of data for each of
the 4 symbols.
This takes 4\(\times\) more space per character, but since the text doubles in space as
well, the overhead only doubles to 6.25\%.
Alternatively, 256 bp (512 bit) blocks can be used to reduce the number of cache
misses, using 12.5\% overhead.

\parag{SPIDER: interleaving bits for minimal cache misses.}
\emph{SPIDER} \cite{spider} has only 3.3\% overhead and reduces the number of cache misses from 2 to (nearly) 1
by interleaving L1 with the bitvector itself (like BWA-MEM), instead of interleaving L1 with L2:
each cache line stores a 16-bit L1 delta, and 496 input bits.
L2 superblocks store a 64-bit offset for each 128 blocks, taking only 0.1\% extra
space and thus likely fitting in a cache.

\parag{Pairing: halving the overhead.}
\emph{Pairing} (pfBV) \cite{engineering-rank} is an idea that halves the memory
overhead again, to 1.6\%. Compared to PastaWide, instead of storing 16-bit (L1)
deltas to the start of \emph{each} 512-bit block,
here we store 16-bit deltas to the middle of each \emph{pair} of 512-bit blocks.
Then, the second block can add a prefix-popcount to this as usual, while
the first block can \emph{subtract} a \emph{suffix}-popcount instead.
Similarly, the 64-bit L2 offset is to the \emph{middle} of a twice-as-large superblock.
This is similar to the \emph{alternate counters} idea for the FM-index
\cite{fm-gpu}, where, for alphabet size 4, each block stores half the offsets.
A small complication with this design is that conditionally popcounting a
prefix \emph{or} suffix of bits is slightly slower. Instead, Gottlieb and Reinert \cite{engineering-rank}
introduce a lookup table that stores a precomputed mask for each position.
Lastly, for \(\sigma=4\), this paper uses the transposed layout of
AWFM, but calls it \emph{scattered} instead.

\pagebreak
\subsection{Further implementations}
\label{implementations}
In anticipation of the evaluations, we list some specific Rust implementations.

\parag{QWT.} \texttt{qwt} (\href{https://github.com/rossanoventurini/qwt}{github:rossanoventurini/qwt}) implements \texttt{RSQVector256} and \texttt{RSQVector512}
corresponding to the Quad Vectors in the paper \cite{quad-wavelet-tree} with
12.5\% and 6.25\% overhead. It further contains \texttt{RSWide}, which implements the
PastaFlat structure \cite{pasta} (omitting the L3 layer), and \texttt{RSNarrow},
which exactly implements Rank9.

\parag{Sux.} \texttt{sux} (\href{https://github.com/vigna/sux-rs}{github:vigna/sux-rs}) \cite{sux-repo} contains an implementation of Rank9, as well as five
versions of \emph{RankSmall}.
These are all variants on Rank9, but use Poppy's 64-bit L3 to allow for 32-bit L2
values. They vary in the number of \texttt{u32} used to store the L1 values and the
width of the L1 values. A special case is \texttt{RankSmall3} (3.125\% overhead), which stores 3 11-bit
values in a single 32-bit word by using 0-extension for the implicit high 0-bit of
the first value.

\parag{Bitm.} \texttt{bitm} (\href{https://github.com/beling/bsuccinct-rs}{github:beling/bsuccinct-rs}) is part of bsuccinct \cite{bsuccinct}. Its \texttt{RankSimple} (6.25\%
overhead) stores a 32-bit L2 offset for every 512 bit block.
\texttt{RankSelect101111} (read: 10-11-11) has 3.125\% overhead and is the same as
\texttt{RankSmall3} of sux.

\parag{Genedex.} \texttt{genedex} (\href{https://github.com/feldroop/genedex}{github:feldroop/genedex}) \cite{genedex} implements variants of the data structures of
\cite{engineering-rank}. It is designed for \(\sigma>2\), but also supports
\(\sigma=2\).
\texttt{Flat512} stores the text using 4 indicator bitvectors and uses 4 interleaved copies of SPIDER, one for each symbol.
\texttt{Flat64} is the same but with 64-character blocks.
\texttt{Condendensed512} implements the flattened bit vectors (fBV) of
\cite{engineering-rank}, with blocks representing 512 transposed
characters, and using \(\sigma\) interleaved copies of PastaWide.

\parag{Further Rust implementations.}
We did not include the following libraries in the evaluations because they are not (close to) Pareto optimal.
 \textbf{\texttt{Bio}} \cite{rust-bio} has a \texttt{RankSelect}
structure that stores a 64-bit offset after every few 32-bit words, but is not
very optimized.
 \textbf{\texttt{RsDict}} \cite{rs-dict} implements a compact
encoding \cite{simple-rank-select}, making it relatively slow.
 \textbf{\texttt{Sucds}} \cite{sucds} implements Rank9, which is already covered.
 \textbf{\texttt{Succinct}} \cite{succinct} provides both Rank9 and
JacobsonRank, which is both slower and larger.
 \textbf{\texttt{Vers\_vecs}} \cite{vers-vecs} implements PastaWide, but with superblocks spanning \(2^{13}\) rather than \(2^{16}\) bits.

\section{BiRank}
\label{sec:org557266e}
Let \(T = t_0\dots t_{n-1}\) be a text of \(n\) binary characters \(\{0,1\}\).
For a query \(q\) (\(0\leq q\leq n\)),
\(\rank(q) = \sum_{i\in \{0,\dots,q-1\}} [t_i = 1]=\sum_{i\in [q]} t_i\) counts
the number of 1-bits in the first \(q\) characters of the text.

BiRank is a data structure that answers \(\rank(q)\) queries in constant time
using 3.28\% space overhead.
It can be constructed in parallel on multiple threads from a slice of already-packed data and
provides \texttt{rank} and \texttt{prefetch} functions.
The description below refers to the lines of the simplified Rust code for
querying QuadRank in \cref{snippets}.

\parag{Single cache-miss queries.}
We aim to minimize the number of cache misses on large (many GB) inputs,
to enable efficient usage in high-throughput settings where the
memory bandwidth is the bottleneck.
A single cache miss is inevitable, and so we must avoid any further cache misses.
This means that any additional data should fit in L3 cache, which is not the
case for non-interleaved layouts.
For example, pairing has 1.6\% overhead, which would only support 1 GiB of input
with a 16 MiB L3 cache.

\parag{Interleaved L1.}
Thus, like SPIDER \cite{spider}, BiRank inlines a 16-bit L1 delta \(b_j\) into
each block/cache line \(j\) (lines 7-8 of \cref{quadrank-code}), so that each of \(\lceil n/B\rceil\) blocks covers \(B:=512-16=496\) bits.
The \(\lceil n/S\rceil\) superblocks cover \(S:=128\cdot B\) bits each, and a second much smaller array
stores a 32-bit L2 offset \(s_i\) for each superblock \(i\).

\parag{Shifted 32-bit L2 offset.}
Poppy \cite{poppy} uses an additional 64-bit third level L3, so that 32-bit
L2 values are sufficient. Even though this L3 layer is already very small, we remove it completely.
Rather than directly encoding \(\rank(i\cdot S)\), the rank at the start of a superblock, we store
$$s_i := \lfloor \rank(i\cdot S)/2^{11}\rfloor$$
in the 32-bit L2
value. The remainder \(\rank(i\cdot S)\bmod 2^{11}\) will be added to the 16-bit
\(b_j\) delta for each block in the superblock.
This configuration supports inputs up to \(2^{43}\) bits, or 1 TiB, since
\(n<2^{43}\) implies \(\rank(i\cdot S)/2^{11} < 2^{32}\).

\parag{Size of a superblock.}
Each superblock must contain at most \(2^{16}\) bits, so that the 16-bit \(b_j\)
can represent their deltas. Thus, we could fit \(\lfloor 2^{16} / B\rfloor = 132\)
blocks inside each superblock, but we round this down for computational
efficiency: \(S := 128\cdot B\).

We also implemented a variant of the pairing of superblocks technique of
\cite{engineering-rank}, which doubles the superblock size \(S\) and halves the
cache usage, see \cref{pairing-math}. We decided not to use it though:
while the reduced cache usage could be beneficial, in practice, the gains are
inconsistent and small at best.

\parag{Overhead.}
The overhead of the block deltas is \(16 / B = 3.226\%\), whereas the
superblocks have an overhead of \(32 / S = 0.05\%\). Thus, the total overhead is
3.28\%, and the superblock array fits in a 16 MiB L3 cache for inputs up to 32 
GiB.

\parag{L1-delta to the middle.}
To reduce the amount of work needed for popcounting, we apply a variant of the pairing
technique \cite{engineering-rank}: the 16-bit L1 value is not the delta from
the start of the superblock to the
\emph{start} of the current block (\(\rank(j\cdot B)\)), but instead to the \emph{middle} of
the current block (\(\rank(j\cdot B + 240)\), after taking into account a 16-bit padding, line 9):
$$ b_j := \rank(j\cdot B + 240) - 2^{11}\cdot s_{\lfloor j/128\rfloor}. $$
By construction, these values are indeed bounded by \(b_j \leq S + (2^{11}-1) < 2^{16}\) (line 8).

\parag{Queries.}
A query for position \(q\) first determines the superblock \(i_q = \lfloor q/S\rfloor\)
and block \(j_q = \lfloor q/B\rfloor\). Then, 
we compute the rank of the middle of the block as \(\rank(j\cdot B+240) = 2^{11}\cdot s_{i_q} + b_{j_q}\).
We then make a case distinction on whether \(q\) lies left or right of the middle
of its block to determine the final value
\begin{equation}
\label{eq:query}
\rank(q) = 2^{11}\cdot s_{i_q} + b_{j_q} + \begin{cases}
\phantom{-}\sum_{k\in \{j_q B+240, \dots, q-1\}} t_i & \textrm{if }q \geq j_q \cdot B + 240,\\
-\sum_{k\in \{q, \dots, j_q B+240-1\}} t_i & \textrm{if }q < j_q \cdot B + 240.
\end{cases}
\end{equation}
In code, we popcount up to 256 bits (lines 15 and 18): either a suffix of the first half, or a prefix of
the second half. The conditional negation (line 20) is optimized to a branchless \texttt{cmov} instruction.

\parag{Masking.}
Instead of a for loop over the 64-bit words in the block and bit-shifting, we
prefer a branchless technique that always covers the full 256 bits.
Uncounted bits are masked out (line 18) via a 256-bit mask that is
precomputed (line 11) for each \(0\leq (q\bmod B)< 512\), again following
\cite{engineering-rank}. These are simply stored as a 16 KiB array \texttt{[u256;
512]}, which fits in a typical 32 KiB L1 cache.

\parag{Prefetching.}
In order to facilitate efficient batching algorithms (see \cref{fm-index}), we provide
a \texttt{prefetch(q)} function that starts loading the two cache lines containing
\(s_{\lfloor q/S\rfloor}\) and \(b_{\lfloor q/B\rfloor}\) that are needed for \(\rank(q)\).
For simplicity and reliability, we prefetch into all levels of the cache hierarchy.

\parag{Parallel construction.}
The parallel construction algorithm used for both BiRank and
QuadRank builds on the \texttt{rayon} crate.
First, we count the number of 1-bits in each superblock, which is trivial to
parallelize. Then, we take a prefix sum over these counts in a (non-parallel) linear
pass, so that we know the number of 1-bits preceding each superblock. This then
allows us to fully construct all superblocks (and their contained blocks) in parallel.

\subsection{Variants}
\label{sec:org6cda2ac}
We consider a few larger but faster variants of BiRank. The ones marked with a *
are the best for each overhead and chosen for the evaluations.

\begin{enumerate}
\item \itemname{BiRank16*} (3.28\% overhead) is the original as described above and inlines a 16-bit value in each
cache line.
\item \itemname{BiRank32} (6.67\%) is identical but stores a 32-bit value instead, doubling the overhead.
This allows for a much (\(\approx 2^{16}\times\)) smaller L2 array.
\item \itemname{BiRank16x2*} (6.72\%) stores \emph{two} 16-bit deltas, to 1/4th and 3/4th into the block.
Then, only a quarter of the cache line (2 64-bit words) has to be popcounted.
\item \itemname{BiRank23\_9} (6.67\%) takes a middle ground: it stores a 23-bit L1 delta to 1/4th of the block,
and a 9-bit ``L0.5'' delta (\(\leq 256\)) from there to 3/4th.
\item \itemname{BiRank64} (14.3\%) directly stores a 64-bit value instead, completely removing the need
for a separate L2 level.
\item \itemname{BiRank32x2*} (14.3\%) doubles the overhead again and stores two 32-bit L1 values, shrinking
the L2 array.
\item \itemname{BiRank64x2*} (33.3\%) \emph{again} doubles the overhead, and completely removes the L2 level.
\item \itemname{BiRankR9} (33.3\%) is an inline version of Rank9: it inlines a 64-bit
L2 offset, followed by a 64-bit word containing 6 9-bit deltas to the start
of each remaining 64-bit word.
\end{enumerate}
\section{QuadRank}
\label{sec:orgf8eec8a}
QuadRank is the extension of BiRank to the 2-bit DNA alphabet.
It can be constructed in parallel on multiple threads from bitpacked data.
Rank queries can be either for a
specific symbol (\(\rankone(q, c)\)), or for all 4 symbols at once (\(\rankall(q)\)).
We do not provide a dedicated function to count a range, as is
commonly used for the FM-index, because the associated branch-misses would hurt
performance, and cache lines are automatically reused anyway.

As with BiRank, QuadRank is optimized for having as few cache misses as possible.
In particular, the data-layout is nearly the same, but with the L2 and L1 data
replicated for each symbol: each cache line contains 4 16-bit deltas \(b_{j,c}\)
and \(B_4=(512-4\cdot 16)/2=224\) characters.
Superblocks cover 256 blocks (\(S_4:=256\cdot B < 2^{16}\)), and for each we store
4 shifted 32-bit offsets
\(s_{i, c} := \lfloor \rankone(i\cdot S_4, c) / 2^{13}\rfloor\).
We now divide by \(2^{13}\), since \(S + (2^{13}-1) < 2^{16}\), which allows inputs
up to \(2^{45}\) characters or 8 TiB.
The overhead over the \(b_i\) deltas is \(4\cdot 16 / (2B) = 14.29\%\) and the
overhead of the offsets is \((4\cdot 32) / (2S) = 0.11\%\), for 14.40\% overhead in total.
A 16 MiB L3 cache can support over 14 GiB (61 Gbp) of input.

To compute the rank of all 4 symbols at once,
relatively more time is spent on popcounting than in the binary case.
Thus, we detail our optimizations to compute all 4 ranks efficiently.

\parag{Transposed layout.}
Compared to the layout for binary input, the main difference is that we now
store the
input data in \emph{transposed} (or \emph{strided}) layout
\cite{awry-optimized-fm-index,engineering-rank} (as opposed to \emph{packed}).
Ignoring the inlined \(b_{j,c}\) deltas for the moment,
the 256 characters in a block are split into 4 groups of 64.
Each group of 64 characters is encoded as two 64-bit values, one consisting of
the \emph{negation} of all low bits, and one of the \emph{negation} all high bits.
The 4 16-bit deltas replace the 64 bits corresponding to the 32
first characters (\cref{quadrank-code}, line 8), as shown in the bottom row of \cref{ranks}.
The positions matching a symbol are then found by and-ing the two
values together (line 18), after possibly negating one or both (lines 13-17).
This layout makes more efficient use of popcount
instructions, since each counted word now contains up to 64 1 bits, compared to
32 with the packed layout.

\parag{rank1.} Computing the rank for a single character is similar to before \hyperref[eq:query]{(1)}:
we retrieve the superblock offset \(s_{i_q}\), multiply it by \(2^{13}\), and then
add the \(b_{j_q,c}\) for the current block and character.
Lastly, we add or remove the count for up to 128 characters in either the first or
second half of the cache line, processed in two chunks of 64 characters.

\parag{4-way popcount.} To return the rank of all 4 symbols, we essentially do the
above method 4 times in parallel in \texttt{u64x4} 256-bit AVX2 SIMD registers. In particular,
we use a single SIMD lane for each symbol (\cref{quadrank-code}).
To popcount the number of 1-bits in each lane,
we use Mula's algorithm \cite{sse_popcount,avx2_popcount}. Essentially,
this splits each byte into two 4-bit nibbles and for each does a
\texttt{\_mm256\_shuffle\_epi8} instruction to do a 16-way lookup returning the
precomputed number of 1 bits in each nibble. It then adds these two values, resulting in per-byte
popcounts, and finally uses the \texttt{\_mm256\_sad\_epu8} instruction to take a
horizontal sum of the 8 bytes in each 64-bit lane. 
We convert the counts to \texttt{u32x4} and then
conditionally negate them using \texttt{\_mm\_sign\_epi32(counts, u32x4::splat(pos-96))},
which multiplies each lane by the sign of \((q\bmod B)-96\) (i.e., -1, 0, or 1).
\subsection{Variants}
\label{sec:orgad5197b}
Again, we consider a number of slightly faster variants that use larger inline values.
Since returning all 4 counts takes more compute, we specifically focus on
methods that reduce the amount of characters to be counted from 128 to 64.
There is more variation here than in the binary case: we can use packed (P) or
transposed layout (T), and we can avoid using the pairing technique
(bidirectional (B) vs forward (F)) to save the small CPU overhead for negating values.
This is just a small selection of possibilities,
and not all implementations were equally optimized. Those marked with a * are
the fastest for each overhead and have been chosen for the evaluations.

\begin{itemize}
\item \itemname{QuadRank16*} (TB, 14.40\% overhead) is as described above and inlines 4 16-bit values
containing the rank to the middle of each block.
\item \itemname{QuadRank32} (TB, 33\%) instead uses 4 32-bit values, making the L2 array much smaller.
\item \itemname{QuadRank24\_8*} (TB, 33\%) leaves space for 3 groups of 64 characters and splits this
into 3 sub-blocks, storing an L1 delta to the end of the first and third
group. This way, only a 64-character popcount remains.
\item \itemname{QuadRank7\_18\_7} (PB, 33\%) uses a normal packed layout. It stores an 18-bit L1 to
the middle of 6 32-character blocks, and two 7-bit ``L0.5'' deltas to 1/6th and
5/6th.
\item \itemname{QuadRank64*} (TB, 100\%) stores 4 64-bit values, as does BWA-MEM, removing the L2 array. This only
leaves space for 128 characters, so each half is now only 64 of them.
\item \itemname{QuadRank32\_8x4} (PF, 100\%) uses packed layout. It stores a 32-bit L1 delta to the
start of the block, and 4 8-bit ``L0.5'' deltas to each 32-character sub-block.
\item \itemname{QuadRank32x2} (PF, 100\%) stores 2 32-bit L1 deltas to the start and halfway point,
and does a forward scan.
\end{itemize}
\section{Results}
\label{sec:org67e9b6e}
Both our implementation of BiRank and QuadRank and the
evaluations can be found at \href{https://github.com/ragnargrootkoerkamp/quadrank}{github:ragnargrootkoerkamp/quadrank}.
All experiments are run on an AVX2 Intel Core i7-10750H Skylake CPU with 6 cores and
hyper-threading enabled. The frequency is pinned to 3.0GHz. Cache sizes are 32
KiB L1 and 256 KiB L2 per core, and 12 MiB shared L3 cache. Main memory is 64
GiB as dual-channel 32 GiB 3200MHz DDR4 sticks, with the memory controller
running at 2933 MHz.

Benchmarks on a 92-core AMD Zen 4 EPYC with 12 DDR5 memory channels can be found
in the appendix and give mostly similar results (\cref{epyc}). The appendix also
contains additional plots analysing the CPU time for very small inputs
(\cref{small-n}), as well as statistics on the number of measured last-level cache
misses per query (\cref{cache-misses}).

We only compare Rust implementations, since our aim is to provide a ready-to-use
Rust library as well. Furthermore, cross-language function calls would
likely prevent the compiler from optimizing all code equally, and
re-implementing comparable benchmarks in C++ and getting all libraries to work
was deemed infeasible.
\subsection{BiRank}
\label{evals-birank}
We compare BiRank and its variants against the Rust crates mentioned in \cref{implementations}.
In order to make the evaluations with prefetching fair, we have created PRs
adding support for this to each of them.\footnote{\href{https://github.com/vigna/sux-rs/pull/98}{github:vigna/sux-rs/pull/98},
\href{https://github.com/rossanoventurini/qwt/pull/6}{github:rossanoventurini/qwt/pull/6},
\href{https://github.com/feldroop/genedex/pull/4}{github:feldroop/genedex/pull/4},
\href{https://github.com/beling/bsuccinct-rs/pull/14}{github:beling/bsuccinct-rs/pull/14}.}

\parag{Benchmark setup.}
For each run, we first build each data structure on a random 4 GiB input (in
parallel, if possible) and generate 10 million random query positions.
Then we run three types of benchmarks. In the first, we measure the average \textbf{latency} of
sequential queries, by making each query dependent on the result of the previous one.
In the second, we measure the \emph{inverse throughput} (i.e., amortized time per
query, which we will just call throughput) when processing random queries in
a \textbf{for loop}: \texttt{for i in 0..Q \{ BiRank::rank(queries[i]) \}}.
We stress here that CPUs can use pipelining and out-of-order execution to execute multiple
(up to at least 4) iterations of the loop in parallel.
Thus, we add a third mode where we explicitly process many items at once and we add
\parag{prefetching}, where we prefetch the required cache lines 32
iterations ahead: \texttt{for i in 0..Q \{ BiRank::prefetch(queries[i+32]); BiRank::rank(queries[i]);
\}}\footnote{In practice, we must also prefetch the upcoming query values themselves.
We keep the memory system so busy that it does not have time to do this by
itself, leading to cache misses on the query values themselves if we do not
prefetch them.}
We then repeat these three benchmarks when running in parallel on 1, 6, and 12
threads, where each thread has its own independent set of 10 M queries.
Each reported measurement is the median of 3 runs.

\parag{Excluded libraries.}
SPIDER (\href{https://github.com/williams-cs/spider}{github:williams-cs/spider}) \cite{spider} was not yet implemented in Rust, so we made a
variant of BiRank that approximately uses SPIDER's linear-scan for
popcounting inside a block. Unfortunately, we were unable to compare the
performance against the original C implementation.
Pairing (\href{https://github.com/seqan/pfBitvectors}{github:seqan/pfBitvectors}) \cite{engineering-rank} also has only
been implemented in C++. Nevertheless, genedex was reported to be faster (personal communication).
Lastly, we exclude the dynamic B-tree (\href{https://github.com/jermp/mutable\_rank\_select}{github:jermp/mutable\_rank\_select})
of \cite{rank-select-mutable-bitmaps}, but consider a Rust reimplementation of
this work a promising direction for future work on select specifically.

\begin{figure}[t]
\centering
\includesvg[width=\linewidth]{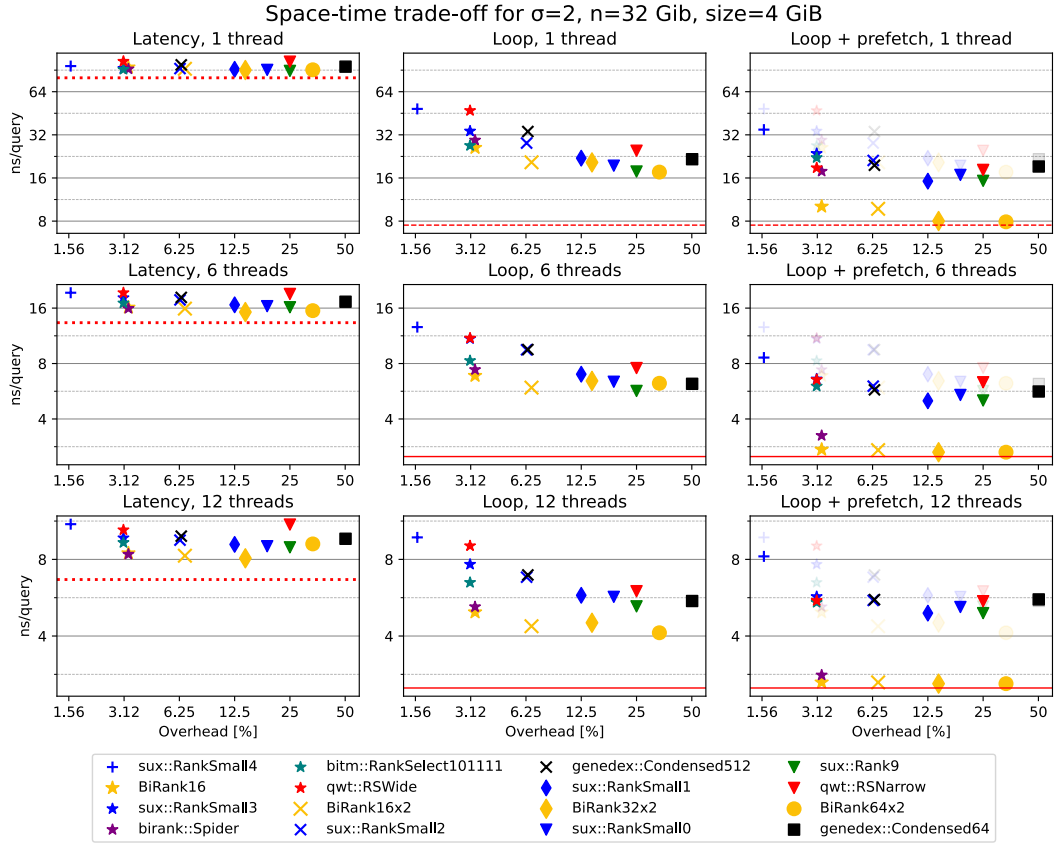}
\caption{\label{birank-plot}Log-log space-time trade-offs for rank structures on binary input of total size 4 GB. The top/middle/bottom row show results for 1/6/12 threads on a CPU with 6 cores. The left/middle/right column show results for the latency, the throughput of a for loop, and the throughput of a for loop with prefetching. Red lines indicate: (left) the roughly 80 ns RAM latency divided by the number of threads, (top mid/right) the 7.5 ns/read maximum random-access RAM throughput of 1 thread, and (rest) the 2.5 ns/cache line total random-access RAM throughput. In the right column, the transparent markers repeat the for-loop throughput. The legend is sorted by increasing overhead.}
\vspace{-1.5em}
\end{figure}

\parag{Lower bounds.}
The results are in \cref{birank-plot}.
There are different lower-bounds on the throughput: the measured latency of the RAM is around 80
ns/read, which gives a lower-bound of 80/\(t\) ns/query for \(t\) threads (dotted red
lines). When using only a single thread, we are further bound by its maximum
random access throughput of around 7.5 ns per cache line (dashed red lines). In
all other cases, we are limited by the 2.5 ns/cache line throughput of the RAM
(solid red lines).

\pagebreak
\parag{Analysis.} 
The first observation is that the latency of all methods is similar, as this is
always limited by the memory latency. Furthermore, processing queries in a loop
is around 4\(\times\) faster with 1 thread, and up to 8\(\times\) with
added prefetching. Using 12 threads halves the gap, but
nevertheless, processing multiple independent queries in each thread
should be preferred to exploit instruction-level parallelism. 

Looking at the middle column, we see that BiRank is just slightly better than other
methods when using a single thread. This grows to 1.4\(\times\) speedup when
using many threads, where it likely benefits from the reduced memory bandwidth.
We see that BiRank16 (the default) is smaller but slightly slower than the larger BiRank
variants. BiRank16x2 has double the overhead and is slightly faster, while the
variants with larger overhead (that shrink/remove the array of superblock offsets) only provide
very minimal gains.

After adding prefetching (right column), we see that all methods improve
compared to the shaded data points without prefetching, most somewhere around 1.5\(\times\). Whereas non-interleaving
methods are limited to around 16 ns/query, BiRank is able to reach the
hard limit of 8 ns/query that each thread needs per cache line.
Here, the larger variants benefit from requiring a bit less compute compared to
the smaller variants.
When using prefetching from multiple threads, the situation is the same:
BiRank can fully exhaust the RAM random-access throughput, even in its smallest
configuration, whereas other methods are 2\(\times\) as slow. We also see that
prefetching speeds up BiRank around 2\(\times\) compared to just a plain for loop.
A special case here is SPIDER, which is also interleaved. With a single thread,
the branch-misses of popcounting in a for loop hurt its performance, but it
becomes as fast as BiRank and memory bound when multithreading.
\subsection{QuadRank}
\label{evals-quadrank}
We now run the same set of experiments to compare QuadRank against QWT and
genedex on size-4 alphabets. On additional feature is that we compare both
\(\rankone(q, c)\) and \(\rankall(q)\). For the other libraries, \(\rankall\) is implemented
naively by simple calling \(\rankone\) four times, whereas QuadRank is primarily
optimized for this case.

\begin{figure}[t]
\centering
\includesvg[width=\linewidth]{./plots/plot-laptop-st-4-large}
\caption{\label{quadrank-plot}Space-time trade-off of rank structures on size 4 alphabet on 4 GiB input. Compared to \cref{birank-plot}, here we benchmark both \(\rankone(q, c)\) (small markers), and \(\rankall\) (large markers).}
\vspace{-1.5em}
\end{figure}

Overall, the situation here is similar to the binary case.
In most settings (single or multithreaded, in a for-loop or with prefetching),
the default 14.4\% overhead version of QuadRank is around 1.4\(\times\) faster than
the 12.5\% overhead version of QWT, and 2\(\times\) faster for \(\rankall\). Also at
large overhead, QuadRank is faster than all genedex variants.
In high-throughput settings, QuadRank can again saturate the memory bandwidth
and answer one query per cache line, being 2\(\times\) faster than all other
methods below 100\% overhead. With prefetching,
QuadRank shows little to no overhead for computing \(\rankall\):
Even with the additional SIMD operations to compute all ranks, it is still memory bound.
\section{Conclusion}
\label{sec:org20f63cb}
We surveyed a large number of existing rank structures
and, inspired by them, developed BiRank and QuadRank.
Their main novelty is in bringing together many independent parts, and applying
them to size-4 alphabets. We benchmarked them in a high-throughput setting,
with many threads and batching of multiple queries inside each thread.

For binary input, the previous best data structure is
SPIDER \cite{spider}.
BiRank is usually slightly faster than our reimplementation of SPIDER.
When multithreading, both
are around 1.5\(\times\) faster than other methods.
Additional prefetching improves all methods, and
doubles the throughput of BiRank, making it 2\(\times\) faster than
all other methods with prefetching.
BiRank benefits from single cache-miss queries, compared to
two for most other methods, and thus makes optimal use of the limited memory bandwidth.
For QuadRank, the improvement over existing methods is
already 1.5\(\times\) without multithreading, and again 2\(\times\) with prefetching.

Using batching and prefetching with our memory-bandwidth-frugal
methods allows up to 3\(\times\) higher throughput than sequential processing, and
this increases to 6\(\times\) speedup on a server with many more cores but higher memory
latency.

These results also translate to up to 4\(\times\) speedups over the
state-of-the-art when used in an FM-index, and we hope that multithreading,
batching, and prefetching become standard in both applications and benchmarks.

In general, we observe that designing data structures for high-throughput, by minimizing
cache misses and branch-misses and adding support for prefetching, can give big
gains when benchmarks are adjusted accordingly.

\parag{Future work.}
Future work remains in generalizing and optimizing the library for other
platforms than AVX2. As the current code was optimized for Skylake (2015),
it is likely that more modern platforms (Golden Cove, 2021 or zen 5, 2024) admit different trade-offs.
For AVX512, there are dedicated popcount instructions that could be used, while
ARM NEON only supports 128-bit instructions and will need further work.

Additional features could be in-place parallel construction and
a \(\mathsf{prefix\_rank}(q, c)\) operation that counts the number of occurrences
of characters \emph{at most} \(c\) to support the bidirectional FM-index.
Lastly, our FM-index could be extended with
support for \(\mathsf{locate}\) queries.
\section{Acknowledgements}
\label{sec:org7434a95}
I thank Heng Li for motivating me to start this project in an attempt to
speed up BWA-MEM. I also thank those who were involved in discussions
surrounding this project or gave feedback on the text:
Rick Beeloo, Piotr Beling, Felix Leander Droop, Simon Gene Gottlieb, Florian Kurpicz,
Rob Patro, Giulio Ermanno Pibiri, and Peter Sanders.

\bibliographystyle{plainurl}
\bibliography{bibliography}

\newpage
\AddToHook{cmd/appendix/before}{%
    \crefalias{section}{appendix}%
    \crefalias{subsection}{appendix}
}
\appendix

\section{Code snippets}
\label{snippets}
%% Simplified code for computing the rank inside a QuadRank block is shown in \cref{quadrank-code}.

\begin{listing}[!h]
  \begin{minted}[linenos]{Rust}
type Block = [[u64; 4]; 2];        // 2 halves of 4 64-bit values.
// Little-endian mask values:
// pos   0: 111...111, pos   1: 011...111, ..., pos 127: 000...001
// pos 128: 000...000, pos 129: 100...000, ..., pos 255: 111...110
static MASKS: [[u64; 2]; 256] = ...;
fn rank1(block: &Block, mut q: u64, c: u8) -> u64 {
  let block_u16: &[u16; 32] = transmute(&block); // Cast block to u16's.
  let delta = block_u16[c + (c&2)] as u64;  // Jump over transposed data.
  q += 32;                         // Adjust for skipping 32 characters.
  let half = block[q/128];         // Read the low or high half of bits.
  let masks: [u64; 2] = MASKS[q];  // Read bitmask for the relevant bits.
  let mut popcount = 0;
  let cl = -(c as u64 & 1);        // Cast 0 or 1 to 0 or u64::MAX.
  let ch = -(c as u64 >> 1);
  for i in 0..2 {                  // Iterate over 2 words in the half.
    let l = half[2*i  ] ^ cl;
    let h = half[2*i+1] ^ ch;      // l&h indicates occurrences of c.
    popcount += (l & h & masks[i]).count_ones();
  }                                // The if is optimized into a cmov.
  if q < 128 { delta - popcount } else { delta + popcount }
}
  \end{minted}
  \caption{\label{quadrank-code}Simplified code for computing the character count of a prefix of a block in QuadRank16.}%
\end{listing}

\section{Pairing superblocks}
\label{pairing-math}
Here we give a variant on the idea of pairing superblocks \cite{engineering-rank}.
Just like we store block offsets \(b_j\) to the middle of a block and then branch
on adding or removing to/from that, we can also let the superblock offsets \(s_i\)
be to the middle of a double-sized superblock. Let \(S' = 2S=256\cdot B\). We
now store \(\lfloor n/S'\rfloor\) 32-bit superblock offsets \(s'_i := \lfloor \rank(i\cdot
S' + S'/2)/2^{11}\rfloor\). Given a block \(j\) in superblock \(i_j := \lfloor
j/256\rfloor\), its delta is incremented by \(S'\) when it is in the lower half of
the superblock:
\begin{equation*}
b'_j := \rank(j\cdot B + 240) - 2^{11}\cdot s'_{i_j} + \begin{cases}
0 & \textrm{if } j \geq 256\cdot i_j + S'/2 \\
(S'/2 - 240) & \textrm{if }j < 256\cdot i_j + S'/2
\end{cases}.
\end{equation*}
For blocks in the upper half of a superblock this is \(< S'/2\) as before.
In the left half, the uncorrected value is the negation of the number of 1 bits
between the block middle \(j\cdot B+240\) and superblock middle \(i_j\cdot S' +
S'/2\), which is between \(0\) and \(S'/2 - 240\). After the correction, we obtain a
non-negative value \(<S'/2\) again.
Queries are modified accordingly to remove this extra term again:
\begin{align*}
  \rank(q) =& 2^{11}\cdot s'_{i_q} + b'_{j_q} -
  \begin{cases}
  0 & \textrm{if } q \geq S'\cdot i_q + S'/2 \\
  S'/2 & \textrm{if }q < S'\cdot i_q + S'/2
  \end{cases}\\
   & +
   \begin{cases}
  \phantom{-}\sum_{k\in \{j_q B+240, \dots, q-1\}} t_i & \textrm{if }q \geq j_q \cdot B + 240\\
  -\sum_{k\in \{q, \dots, j_q B+240-1\}} t_i & \textrm{if }q < j_q \cdot B + 240
  \end{cases}.
\end{align*}
\section{Additional results}
\label{sec:org79246be}
\subsection{Cache misses per query}
\label{cache-misses}
\cref{cache-misses-2} and \cref{cache-misses-4} contain experimental measurements of the
number of last-level cache misses each method has on an input of size 4GiB,
that is, the average number of cache lines fetched from RAM per query.
As expected, BiRank and Quadrank have 1 cache miss per query, while nearly all other
methods require at least 2 cache misses.

\begin{table}[htbp]
\caption{\label{cache-misses-2}Average number of last-level cache misses per query on an input of 4 GiB, $\sigma=2$.}
\centering
\begin{tabular}{lrr}
Ranker & Space overhead (\%) & Cache misses per \(\rank\)\\
\hline
\textsf{sux::Rank9} & 25.0 & 2.05\\
\textsf{sux::RankSmall0} & 18.8 & 2.17\\
\textsf{sux::RankSmall1} & 12.5 & 2.03\\
\textsf{sux::RankSmall2} & 6.3 & 2.25\\
\textsf{sux::RankSmall3} & 3.1 & 2.11\\
\textsf{sux::RankSmall4} & 1.6 & 2.31\\
\textsf{qwt::RSNarrow} & 25.0 & 2.05\\
\textsf{qwt::RSWide} & 3.1 & 1.99\\
\textsf{genedex::Condensed64} & 50.2 & 3.73\\
\textsf{genedex::Condensed512} & 6.4 & 3.27\\
\textsf{bitm::RankSelect101111} & 3.1 & 2.25\\
\textsf{birank::Spider} & 3.3 & \textbf{1.02}\\
\textsf{BiRank64x2} & 33.3 & \textbf{1.01}\\
\textsf{BiRank32x2} & 14.3 & \textbf{1.01}\\
\textsf{BiRank16x2} & 6.7 & \textbf{1.02}\\
\textsf{BiRank16} & 3.3 & \textbf{1.02}\\
\hline
\end{tabular}
%% \end{table}
\vspace{1em}
%% \begin{table}[htbp]
\caption{\label{cache-misses-4}Average number of last-level cache misses per query on an input of 4 GiB, $\sigma=4$.}
\centering
\begin{tabular}{lrrr}
Ranker & Space overhead (\%) & Cache misses per \(\rank\) & per \(\rankall\)\\
\hline
\textsf{genedex::Flat64} & 166.9 & \textbf{1.38} & 2.02\\
\textsf{genedex::Flat512} & 106.6 & \textbf{1.35} & 2.32\\
\textsf{genedex::Condensed64} & 50.2 & 2.46 & 2.58\\
\textsf{genedex::Condensed512} & 6.4 & 3.34 & 3.41\\
\textsf{qwt::RSQ256} & 12.5 & 2.01 & 2.00\\
\textsf{qwt::RSQ512} & 6.3 & 2.22 & 2.35\\
\textsf{QuadRank64} & 100.0 & \textbf{1.01} & \textbf{1.01}\\
\textsf{QuadRank24\_8} & 33.3 & \textbf{1.01} & \textbf{1.00}\\
\textsf{QuadRank16} & 14.4 & \textbf{1.05} & \textbf{1.05}\\
\hline
\end{tabular}
\end{table}
\subsection{Throughput for small inputs}
\label{small-n}
In \cref{birank-small} and \cref{quadrank-small} we benchmark on small 128 KiB inputs that fit
comfortably in the L2 cache. This way, experiments will be mostly CPU-bound, and
we get an idea of the maximum performance of each method and their relative
computational cost.

As expected, we see a space-time tradeoff, with methods that need more space
typically being faster. Rank9 (25\% overhead) is the fastest, while RankSmall0 is
small but slow. The BiRank variants are all roughly equally as fast, with the
BiRank16 variant (3.28\% overhead) being slightly slower, but still faster
than other methods. 
The server (see next section) is up to 2\(\times\) faster, and surprisingly, even the smallest BiRank
variant is very fast.

For QuadRank, we see that the laptop is faster with \(\rankone\)
than \(\rankall\) queries, while, again surprisingly, the server is faster for
\(\rankall\) queries.

\begin{figure}[htbp]
\centering
\includesvg[width=\linewidth]{./plots/plot-st-2-small}
\caption{\label{birank-small}Space-time trade-off plot of the inverse throughput of rank queries in a for loop on a small 128 KiB binary input that fits in L2 cache.}

\includesvg[width=\linewidth]{./plots/plot-st-4-small}
\caption{\label{quadrank-small}Space-time trade-off plot of the inverse throughput of rank queries in a for loop on a small 128 KiB input over alphabet size 4 that fits in L2 cache. Small markers indicate the time for a \(\rankone\) query that counts only one symbol, while large markers always return all four ranks for \(\rankall\).}
\end{figure}

\subsection{AMD EPYC evals}
\label{epyc}
We replicate the experiments on a large AMD Zen 4 EPYC 9684X server with 96 cores, 192
threads, 96 MiB of L3 cache for each 8 cores, and 12-channel DDR5 RAM.
It has a base clock frequency of 2.55 GHz, but during our experiments, it ran
consistently at 3.7 GHz with 1 thread, at 3.4 GHz when using all 192 threads,
and at 3.0 GHz when using 192 threads with batch processing.
We run experiments using 1, 48, and 96 threads.
Results for 192 threads are nearly identical to using 96 threads.

\begin{figure}[htb]
\centering
\includesvg[width=\linewidth]{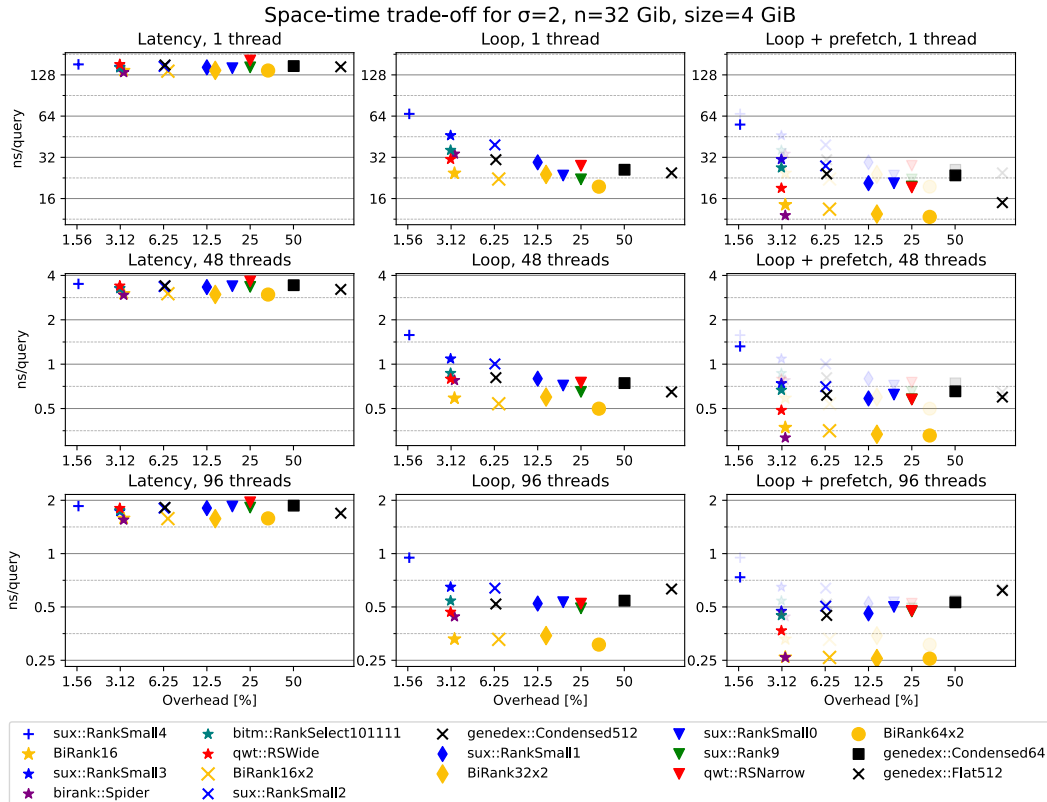}
\caption{\label{birank-scaling-epyc}Scaling with input size for size 2 alphabet.}
\end{figure}

\begin{figure}[ht]
\centering
\includesvg[width=\linewidth]{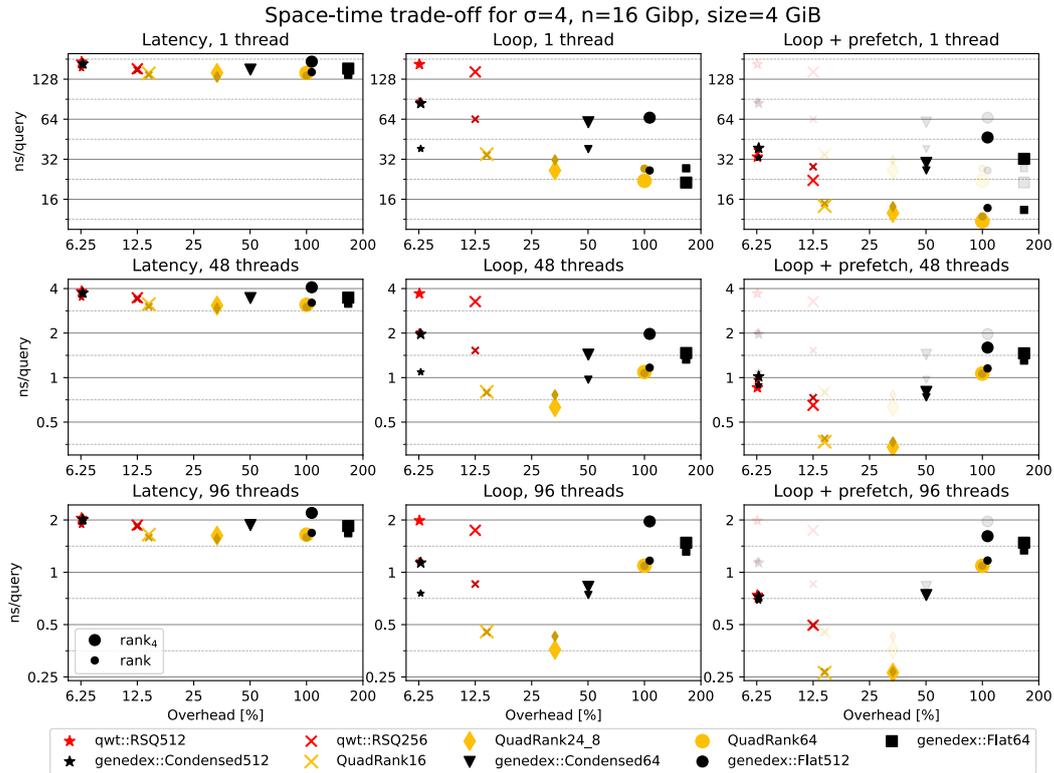}
\caption{\label{quadrank-scaling-epyc}Scaling with input size for size 4 alphabet.}
\end{figure}

\cref{birank-scaling-epyc} and \cref{quadrank-scaling-epyc} show the results, which follow a
similar trend as the earlier results: BiRank and QuadRank are consistently
faster, with the improvement of BiRank becoming more pronounced as we add more threads.
With prefetching, BiRank saturates the memory bandwidth when using 92 threads,
and is then 2\(\times\) faster than nearly all other methods. One notable difference with the
laptop benchmarks is that here, SPIDER is as fast as BiRank when using
prefetching, suggesting that on zen 4, the associated branch misses do not cost nearly as much.
A further difference is that benefit of independent queries over sequential
queries is up to 4\(\times\) here, compared to at most 2\(\times\) on the laptop.
Likely, this is due to the server having a 60\% higher RAM latency
(130 ns vs 80 ns), as well as it having a larger reorder window that can
process more loop iterations in parallel.
At the same time, this reduces the impact of prefetching from 2\(\times\) to 1.5\(\times\).

For \(\sigma=4\), we see that larger data structures are often \emph{slower}.
Likely, this is because the 1.1GiB L3 cache can hold a larger
fraction of the data when the overhead is small.


\section{QuadFm: A Batching FM-index}
\label{fm-index}
To showcase an application of our high-throughput data structure, we develop a
toy implementation QuadFm of an FM-index for the \(\sigma=4\) DNA alphabet.
Inspired by Movi \cite{movi,movi2} and genedex \cite{genedex}
we process queries in batches and prefetch memory for upcoming rank queries.
For simplicity, our implementation only counts the number of matches and does
not support locating them. It only supports exact forward searching, and does not
implement bidirectional search or search schemes
\cite{search-schemes,columba,search-schemes-sahara}, nor in-text verification.
We use a prefix lookup table for the first 8
characters, and handle a single sentinel character (\texttt{\$}) by storing its position
in the Burrows-Wheeler Transform (BWT) \cite{bwa-mem,fm-gpu,bwt}.

The main function \texttt{query\_batch} takes a batch of \(32\) queries and returns for
each the BWT-interval where each query matches.
Similar to genedex, during the processing, we keep an array of the indices of \emph{active} queries whose interval is not
empty yet. As long as there are active queries, we loop over those queries
twice. First, we detect queries that were completed and \emph{swap-pop} them from the
active list, and then prefetch the memory needed for the rank queries. In
a second loop, we perform the rank queries and LF-mapping for each
active query. We do \emph{not} optimize pairs of rank queries for small ranges, to
avoid branch-misses.
\pagebreak
\subsection{Results}
\label{sec:orgf798ef4}
We test our FM-index implementation by building it for a 3.1 Gbp human genome \cite{chm13v2}.
We simulated 500\,000 150 bp reads and applied 1\% uniform random substitution
errors to them. We then count the number of occurrences of each read in both forward and reverse-complement direction.

We compare the size and query throughput of QuadFm against AWRY
\cite{awry-optimized-fm-index} and Genedex \cite{genedex}, which
are both configured to also use an 8-character prefix lookup table and a large suffix-array
sampling factor. In particular, Genedex already supports query batching.
We instantiate QuadRank with each of the rank structures for \(\sigma=4\).
As before, we benchmark on 1, 6, and 12 threads on the laptop CPU, and test
in three modes: \emph{sequential}, where queries are done 1-by-1, \emph{batch}, which does
32 queries in parallel, and \emph{batch+prefetch}, which additionally prefetches
cache lines for the next iteration over all queries.

\begin{figure}[htbp]
\centering
\includesvg[width=\linewidth]{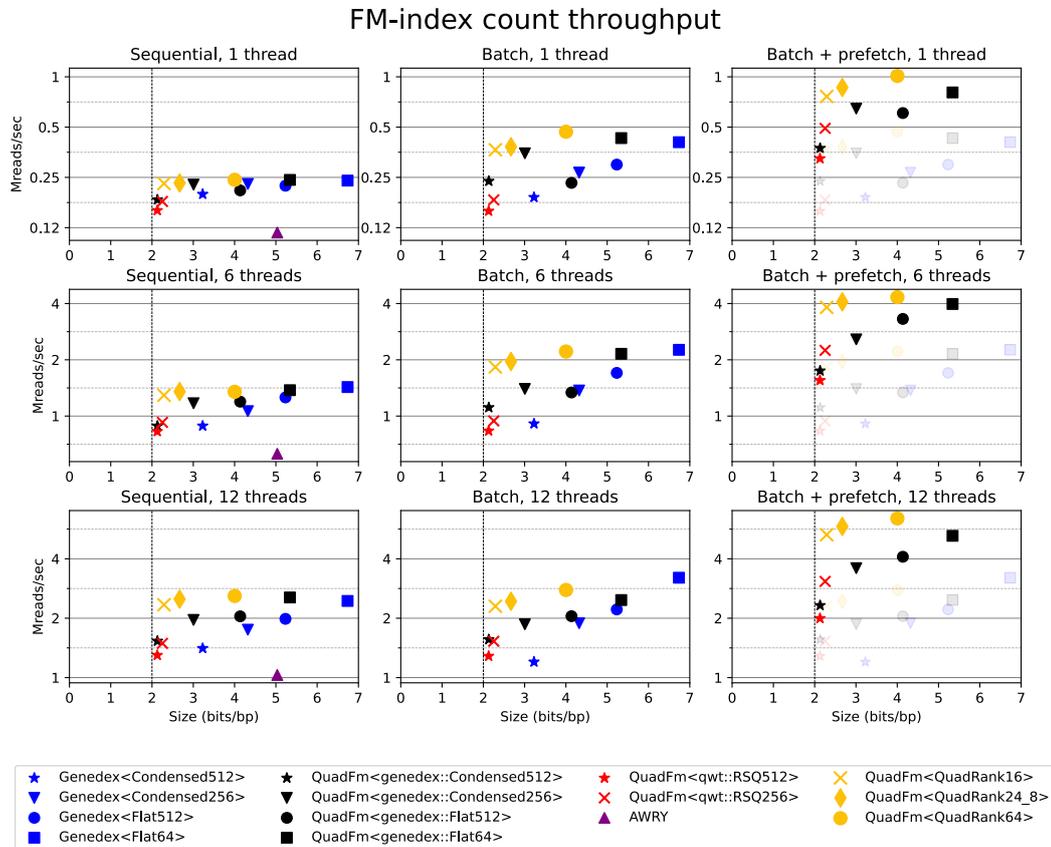}
\caption{\label{fm-evals}Size and throughput of counting exact matches of 150 simulated queries with a 1\% error rate in an FM-index on a human genome. Vertical grey lines indicate the 2 bits/bp lower bound.}
\end{figure}

In \cref{fm-evals}, we see that genedex (blue) is faster than AWRY (purple). Genedex is
consistently around 1 bit/bp larger than using the same rank structures in
QuadFm (black), since it uses a size-5 alphabet to handle the sentinel, but otherwise
they are comparable in performance. Using QuadRank makes QuadFm
both smaller and up to 40\% faster in multithreaded settings.
Prefetching consistently doubles the throughput, and with 12 threads, QuadFm with
QuadRank16 (2.29 bits/bp) is over 4\(\times\)
faster than Genedex' smallest variant (3.2 bits/bp) and 1.65\(\times\) faster
than its fastest variant (6.75 bits/bp).

\end{document}